\documentclass[pdflatex,sn-mathphys-num]{sn-jnl}

\usepackage{graphicx}%
\usepackage{multirow}%
\usepackage{amsmath,amssymb,amsfonts}%
\usepackage{amsthm}%
\usepackage{mathrsfs}%
\usepackage[title]{appendix}%
\usepackage{xcolor}%
\usepackage{textcomp}%
\usepackage{manyfoot}%
\usepackage{booktabs}%
\usepackage{algorithm}%
\usepackage{algorithmicx}%
\usepackage{algpseudocode}%
\usepackage{listings}%
\usepackage{booktabs}

\usepackage{anyfontsize}
\usepackage{xcolor}

\theoremstyle{thmstyleone}%
\theoremstyle{thmstyletwo}%

\theoremstyle{thmstylethree}%

\begin{document}

\title[Article Title]{Stability of Equilibrium Points in Modified Elliptic Restricted Three-Body Problem with Various Perturbation Sources}


\author[1,2]{\fnm{Muhammad Bayu} \sur{Saputra}}

\author*[1]{\fnm{Handhika Satrio} \sur{Ramadhan}}\email{hramad@sci.ui.ac.id}
\equalcont{These authors contributed equally to this work.}

\author[3,2]{\fnm{Ibnu} \sur{Nurul Huda}}
\equalcont{These authors contributed equally to this work.}

\author[4]{\fnm{Leonardus Brahmantyo} \sur{Putra}}
\equalcont{These authors contributed equally to this work.}

\affil*[1]{\orgdiv{Departemen Fisika, FMIPA}, \orgname{Universitas Indonesia}, \orgaddress{\city{Depok}, \postcode{16424}, \country{Indonesia}}}

\affil[2]{\orgdiv{Research Center for Computing}, \orgname{National Research and Innovation Agency}, \orgaddress{\city{Bandung}, \postcode{40135}, \country{Indonesia}}}

\affil[3]{\orgdiv{School of Astronomy and Space Science, Key Laboratory of Modern Astronomy and Astrophysics (Ministry of Education)}, \orgname{Nanjing University}, \orgaddress{\city{Nanjing}, \postcode{210023}, \country{PR China}}}

\affil[4]{\orgdiv{Mathematical Institute}, \orgname{University of Oxford}, \orgaddress{\city{Oxford}, \postcode{OX2 6GG}, \country{United Kingdom}}}


\abstract{This study examines the dynamics of the third body in an elliptic restricted three-body problem (ERTBP) framework, taking into account perturbations from radiation pressure, oblateness, and elongation of the primary bodies, as well as disk-like structures. The objectives are to determine the positions and stability of the equilibrium points, assess how these points shift under the influence of perturbations, and evaluate the dependence of their stability on the orbital eccentricity and perturbation parameters. The ERTBP model is modified to include a radiating, oblate primary body and an elongated secondary body modeled as a finite straight segment, alongside perturbations from a surrounding disk. The system’s equations of motion are numerically solved using parameters from perturbed and classical cases. Equilibrium positions are computed over a range of eccentricities and perturbation values, and stability is analyzed using linearized equations and eigenvalue methods. In all cases, we have found three collinear ($L_1$, $L_2$, $L_3$) and two non-collinear ($L_4$, $L_5$) equilibrium points solutions. The inclusion of radiation, oblateness, elongation using a finite straight segment, and disk perturbation systematically displaces each equilibrium point from its classical location, with the magnitude and direction of the displacement varying with the perturbation parameter. Stability analysis confirms that the collinear points remain linearly unstable under all tested conditions. Meanwhile, non-collinear points are stable under a specific condition. We investigate the stability boundary of these points as a function of orbital eccentricity and we found there is a critical range of eccentricity values within which stability is preserved.}




\maketitle

\section{Introduction}\label{sec1}

The three-body problem is one of the fundamental topics in celestial mechanics that explores the dynamics of three interacting bodies governed by the gravitational, or general central force, law. So far, there has been no general solution to this problem,  although many particular solutions (with various symmetries or restrictions) have been found. Poincaré proved that, in general, the behavior of a three-body system is unpredictable \cite{poincare1892les}. One of the most extensively studied special cases is the {\it restricted three-body problem} (RTBP), where two primary bodies orbit their common center of mass while exerting gravitational forces on a third infinitesimally small body. In the classical formulation, the primaries are typically modeled as point masses or perfect spheres, and only their mutual gravitational interactions are considered \cite{murray1999solar}. In this specific case, the solution where the three bodies lie in a collinear configuration was obtained by Euler, while the equilateral triangle solution was discovered by Lagrange~\cite{szebehely1967theory, Valtonen2006the, mathuna2008int}. 

If the primaries move in circular orbits, the system is known as the circular restricted three-body problem (CRTBP); if they follow elliptical trajectories, it is referred to as the elliptic restricted three-body problem (ERTBP). Both the classical CRTBP and ERTBP yield five distinct equilibrium points, known as Lagrange points. Three of these points ($L_1$, $L_2$, and $L_3$) are collinear and inherently unstable for any values of the mass parameter and eccentricity. The remaining two, $L_4$ and $L_5$, form the vertices of equilateral triangles and remain stable for $0<\mu<\mu_c$, where $\mu$ is the mass ratio and $\mu_c$ is the critical mass parameter \citep{szebehely1967theory}.

The classical RTBP has been modified over time to incorporate additional forces beyond pure gravitation, enhancing its physical realism. One such modification is the inclusion of the radiation pressure as a {\it photogravitational} effect. This force arises from the radiation pressure exerted by luminous celestial bodies, such as stars, on surrounding objects. Therefore, despite its relatively small magnitude, the photogravitational force cannot be neglected when developing more comprehensive mathematical models in celestial mechanics. It was first explored by \citet{radzievskii1950restricted} and expanded by \citet{chernikov1970photogravitational}. Several studies have incorporated the photogravitational effect as an additional perturbative force within the framework of the restricted three-body problem \citep[see, e.g.,][]{markellos1992linear, elipe1997periodic, kushvah2008linear, shankaran2011out, chakraborty2018effect, moneer2023revealing}. In recent work, \citet{lhotka2015effect} and \citet{idrisi2020study} examined the effect of Poynting-Robertson drag and albedo as a photogravitational force, respectively, on the stability of equilibrium points in the ERTBP framework.

In the classical case of the restricted three-body problem, celestial bodies are often assumed to be perfect spheres. However, many astronomical objects are more accurately represented as oblate spheroids. For instance, Earth, Jupiter, and Saturn, as well as stars like the Sun, exhibit some deviations from sphericity that introduce substantial gravitational perturbations. Early attempts to model the influence of an oblate primary on the dynamics of the restricted three-body problem were carried out by \citet{sharma1975collinear,sharma1979effect} and \citet{bhatnagar1977effect}. To date, numerous studies have been conducted to incorporate the oblateness effect as a perturbative force on RTBP \citep[see, e.g.,][]{sharma1990periodic,markellos1996non,douskos2006out,safiya2012oblateness,abd2015stability,moneer2024revealing}.

In addition to considering the oblateness and radiation of the primaries, previous studies have developed mathematical models to assess the influence of a potential disk-like structure surrounding the three-body system \citep[see, e.g.,][]{yousuf2019effects,abozaid2020periodic,mahato2022dynamics,huda2023studying}. The effects of such a disk-like structure have been incorporated into the ERTBP model by \citet{chakraborty2021elliptic}, where the primaries have been treated as point masses. \citet{mia2023analysis} have analyzed the effect of a disk-like structure, oblateness, and radiation pressure on the ERTBP.

Meanwhile, our solar system hosts a wide variety of celestial objects, including those with elongated shapes, such as comets, asteroids, and even dwarf planets. Previous studies have proposed modeling this kind of object as a finite straight segment \citep[see, e.g.,][]{riaguas1999periodic,riaguas2001non,nurul2025model}. This finite straight segment model has been implemented in the study of RTBP by various authors \citep[see, e.g.,][]{jain2014stability,kaur2020effect,verma2023perturbed}.

Previously, \citet{huda2023studying} have analyzed the stability of equilibrium points in CRTBP by incorporating multiple perturbative sources, i.e., radiation pressure, the non-spherical shape of the primary bodies, the influence of a disk-like structure, and the finite straight segment. Here, we extend this study by applying the system to the ERTBP framework. We aim to identify the collinear and non-collinear equilibrium points and examine the stability of the motion of the infinitesimal body. The results were then applied to a perturbed system and compared to the classical case.

The present paper is divided into the following sections. We present our models in Section \ref{sect:model}. The locations of equilibrium points are presented in Section \ref{sect:location}. Section \ref{sect:stability} describes the stability of the system. The conclusion is given in Section \ref{sect:conclusion}.

\section{Formulation of the System}
\label{sect:model}

\begin{figure}
    \centering
    \includegraphics[width=0.75\linewidth]{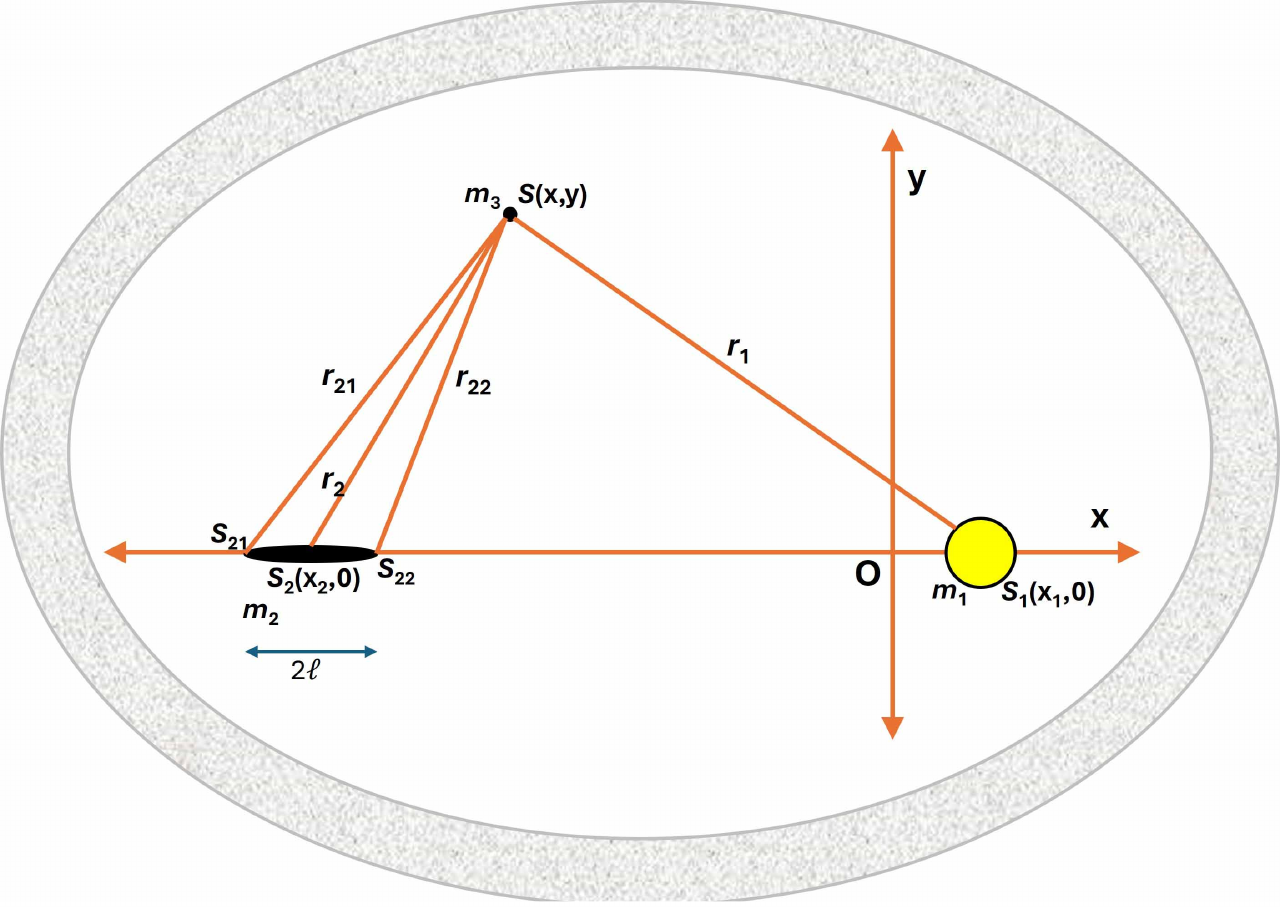}
    \caption{Illustration of the ERTBP system used in this work.}
    \label{fig:system}
\end{figure}

In this work, we analyze a system in which an infinitesimal mass travels under the effect of a larger primary with mass $m_1$ and a smaller primary with mass $m_2$. These primaries orbit each other around their center of mass in an elliptical orbit. The mass parameter of the system is denoted by $\mu = m_2/(m_1 + m_2)$ where $m_2 = \mu$ and $m_1=1-\mu$. The illustration of the system is shown in Fig. \ref{fig:system}.

We consider the larger primary to be an oblate spheroid and a source of radiation. The oblateness parameter is represented by $A = ({AE}^2 - {AP}^2)/{5{R}^2}$ where $AE$ and $AP$ are the equatorial radius and polar radius of the primary body, respectively, and $R$ is the distance between the primaries \citep{sharma1976stationary, sharma1987linear}. The radiation pressure force ($F_p$) works opposite to the gravitational attraction force ($F_g$) and varies with distance according to the same rule. Hence, the total force of the larger primary is $F_g - F_p = F_g(1-(F_p/F_g)) = qF_g$ \citep{chernikov1970photogravitational, kunitsyn1978stability, sharma1987linear} with $q$ is the photogravitational effect parameter with $0<1-q\ll1$. 

The smaller primary is considered elongated and is approximated with a finite straight segment approach with a length of $2\ell$. From \citet{riaguas1999periodic} and \citet{mahato2022dynamics}, the potential of this model is given by 
\begin{equation}
V_f = -\frac{m_2}{2\ell}\log{\left[\frac{r_{21} + r_{22} + 2\ell}{r_{21} + r_{22} - 2\ell}\right]}.    
\end{equation}
In addition to the perturbation effects on the primary bodies, the influence of a disk-like structure that surrounds the system is also taken into account using the potential function of \citet{miyamoto1975three}. By limiting the condition to the $xy$ plane (where $z=0$), the potential function of the disk-like structure ($V_b$) has the form of 
\begin{equation}
    V_b = \frac{M_b}{\sqrt{(R^2 + T^2)}}, 
\end{equation}
where $M_b$ is the total mass of the disk-like structure, $R$ denotes the radial distance of the infinitesimal body defined by $R^2 = x^2 + y^2$, and $T=a+b$. Parameters $a$ and $b$ determine the density profile for the disk-like structure, with $a$ serving as a flatness parameter that determines the flatness of the profile and $b$ as the core parameter that determines the size of the core in the density profile \citep{abozaid2020periodic}.

In a planar elliptic restricted three-body problem, the equations of motion for the third body in dimensionless rotating coordinates are as follows \citep{szebehely1967theory}:
\begin{equation}
\begin{aligned}
\label{eq:motion_eq}
    \frac{d^2x}{d\!f^2} - 2\,\frac{dy}{d\!f} &= \frac{\partial{W}}{\partial{x}}, \\
    \frac{d^2y}{d\!f^2} + 2\,\frac{dx}{d\!f} &= \frac{\partial{W}}{\partial{y}},
\end{aligned}
\end{equation}
with $f$ (true anomaly) as an independent variable. Both primary bodies are positioned along the $x$-axis, with the separation between them normalized to unity. The unit system is normalized so that the gravitational constant is set to unity. The coordinates of the larger primary, the smaller primary, and the third body are given by ($\mu$, $0$), ($\mu-1$, $0$), and ($x$, $y$), respectively. The right-hand side of Eq. (\ref{eq:motion_eq}) represents the first derivative of the potential $W$ and denotes the combined potential resulting from the perturbation sources considered, with
\begin{equation}
    \label{eq:W_init}
    W=\frac{V}{1+e \cos{f}}
\end{equation}
and the function $V$ is the sum of the potential from the centrifugal term, the gravitational terms of the larger primary (including radiation pressure factor and oblateness), the disk-like contribution with mass $M_b$, and the finite straight segment term of the smaller primary with
\begin{equation}
    \label{eq:potential_V}
    V = \frac{1}{2}\left(x^2 + y^2 \right) + \frac{1}{n^2} \Bigg( \frac{m_1 q}{2} \left(\frac{A_1}{{r_1}^3} + \frac{2}{r_1} \right) + \frac{M_b}{\sqrt{R^2 + T^2}} + \frac{m_2}{2\ell}\,\log\left( \frac{r_{21} + r_{22} + 2\ell}{r_{21} + r_{22} - 2\ell} \right) \Bigg).     
\end{equation}

To analyze the long-term behavior of the system, we employ the averaging method described by \citet{grebenikov1964stability} to eliminate the explicit dependence on the true anomaly $f$. The averaged potential, denoted as $\langle W \rangle$, is obtained by integrating the potential $W$ over one complete period of the true anomaly ($f \in [0, 2\pi]$):
\begin{equation}
    \langle W \rangle = \frac{1}{2\pi} \int_{0}^{2\pi} W \, df = \frac{1}{2\pi} \int_{0}^{2\pi} \frac{V}{1 + e \cos f} \, df.
\end{equation}
Since the function $V$ (Eq. \ref{eq:potential_V}) depends only on the pulsating coordinates $(x, y)$ and not explicitly on $f$, it can be treated as a constant during this integration. The remaining integral is a standard definite integral of the form
\begin{equation}
    \int_{0}^{2\pi} \frac{df}{1 + e \cos f} = \frac{2\pi}{\sqrt{1 - e^2}}, \quad \text{for } |e| < 1.
\end{equation}
Substituting this result back into the expression for the averaged potential yields
\begin{equation}
    \langle W \rangle = \frac{V}{2\pi} \left( \frac{2\pi}{\sqrt{1 - e^2}} \right) = \frac{V}{\sqrt{1 - e^2}}.
\end{equation}
Therefore, the Eq. (\ref{eq:W_init}) becomes
\begin{equation}
\begin{split}
    \label{eq:potential_W}
    W = &\frac{1}{\sqrt{1-e^2}} \Bigg[ \frac{1}{2}\left(x^2 + y^2 \right) + \frac{1}{n^2} \Bigg( \frac{m_1 q}{2} \left(\frac{A_1}{{r_1}^3} + \frac{2}{r_1} \right) \\
    &\; + \frac{M_b}{\sqrt{R^2 + T^2}} + \frac{m_2}{2\ell}\,\log\left( \frac{r_{21} + r_{22} + 2\ell}{r_{21} + r_{22} - 2\ell} \right) \Bigg) \Bigg].     
\end{split}
\end{equation}
The distance from the third body to the larger primary body is ${r_1}=\sqrt{(x-\mu)^2 + y^2}$ and the distances of the third body from both of the edge of smaller primary body are ${r_{21}}=\sqrt{(x-\mu+1-\ell)^2 + y^2}$ and ${r_{22}}=\sqrt{(x-\mu+1+\ell)^2 + y^2}$. 

\subsection{Mean motion}

The mean motion $n$ represents the angular velocity at which the primaries revolve around their common center of mass. In RTBP, deriving the mean motion is based on the mean distance variable. The concept of mean distance plays a fundamental role in ERTBP, as it directly influences the mean motion of two primaries. The definition of mean distance is not always straightforward, as different averaging methods can yield different interpretations. 

\citet{Stein1977} discusses several ways to define mean distance, including averaging with respect to the true anomaly, time, and arc length. Mean distance with respect to angle $f$ (true anomaly) results to $a\sqrt{1-e^2}$ (equal to semi-minor axis $b$). This averaging method had been used by \citet{abd2015stability}, \citet{radwan2021location}, and \citet{saputra2024location}. Mean distance with respect to duration of revolution $t$ yields $a(1+e^2/2)$, which is greater than $a$. The mean distance with respect to the arc length of the orbit equals exactly $a$. Several studies have used this kind of approach to calculate the mean motion \citep[see e.g.,][]{narayan2014stability, usha2014effects, huda2015locations, dermawan2015on, duggad2021effects}. In these works, the mean motion is determined by averaging the two-body motion of the primaries within the rotating–pulsating formulation of the ERTBP, after incorporating the specific perturbations included in each model (such as radiation, oblateness, or triaxiality). Although the physical assumptions differ among these studies, they all follow the same conceptual method: the mean motion is defined to be dynamically consistent with the modified gravitational potential and the averaged motion of the primaries. This averaging method yields the same results as using the eccentric anomaly, as in \citet{umar2021impacts}. 
Among these methods, the arc length averaging approach aligns with Kepler’s original definition, which identifies the semi-major axis as the mean distance of the orbit. Despite these variations, using the semi-major axis as the mean distance remains the most consistent approach, providing a reliable framework for analyzing the mean motion and orbital stability in ERTBP.

In this work, we used the mean motion expression derived from the mean distance definition based on arc length, which equals the semi-major axis $a$. The mean motion of the system is determined based on the formulations presented in \citet{sharma2020perturbed} for the oblateness and \citet{aishetu2021perturbation, mahato2022dynamics} for the finite straight segment and disk-like structure effects. Following the framework established by \citet{sharma2020perturbed}, the mean motion for oblateness is derived by analytically integrating the secular perturbation effects of the more massive primary's oblateness ($A$) on the mean anomaly, the argument of periapsis, and the right ascension of the ascending node. These effects are averaged over one complete orbital revolution using analytical integration to define the mean motion. This ensures that the model remains consistent with the laws of orbital motion for bounded elliptic orbits. By extending this methodology to include the additional gravitational perturbations from a finite straight segment (characterized by length $\ell$) and a disk-like structure (characterized by mass $M_b$, core radius $r_c$, and flatness and core parameter $T$), we arrive at a comprehensive expression for the mean motion. Utilizing the methodologies outlined in those references, where the individual contributions from each perturbation are summed to maintain a consistent dynamical model, the expression for $n^2$ is derived as follows. The balance between the centrifugal force and the gravitational force requires \citep{szebehely1967theory,abd2015stability,idrisi2020study,yousuf2019effects,huda2023studying} 

\begin{align} 
\begin{split}
\label{eq:meanmotion0}
m_1 d_1 n^2 = \frac{G m_1 m_2}{d^2-l^2}\left(1+\frac{3A}{a^2\left(1-e^2\right)^2}\left(1+\sqrt{1-e^2}\right)\right)+\frac{G M_b m_1 r_c}{(r^2_c+T^2)^{3/2}}, \\
m_2 d_2 n^2 = \frac{G m_1 m_2}{d^2-l^2}\left(1+\frac{3A}{a^2\left(1-e^2\right)^2}\left(1+\sqrt{1-e^2}\right)\right)+\frac{G M_b m_2 r_c}{(r^2_c+T^2)^{3/2}},
\end{split}
\end{align}
with $d_1$ and $d_2$ are the distances of primaries to the center of mass. To extract the mean motion of the system, we divide the first equation in Eq. (\ref{eq:meanmotion0}) by $m_1$ and the second by $m_2$. Adding the two resulting expressions and using the relation $d = d_1 + d_2$, we obtain
\begin{equation}
    \label{eq:meanmotion1}
	n^2 d = \frac{G(m_1+m_2)}{d^2-\ell^2} \bigg(1+\frac{3A}{a^2\left(1-e^2\right)^2}\left(1+\sqrt{1-e^2}\right)\bigg) + \frac{2G M_b r_c}{\big({r_c}^2 + T^2\big)^{3/2}},
\end{equation}
with $d$ as the mean distance of the primaries equal to the semi-major axis $a$ and normalized to unity ($d=1$). Setting $G=1$, total mass $m_1 + m_2 = 1$, and treating the perturbation parameters as small ($\ell \ll 1$ and $A \ll 1$), the factor $1/{\left(1-\ell^2\right)}$ in Eq. (\ref{eq:meanmotion1}) may be expanded using the binomial series as $1 + \ell^2 + \mathcal{O}(\ell^4)$. In this derivation, we retain terms up to second order in the length parameter ($\ell^2$) and first order in the oblateness ($A$), while neglecting higher-order contributions such as $\mathcal{O}(\ell^4)$ and the mixed term $A\ell^2$. This approximation is consistent with the numerical values adopted later in this work. Therefore, the mean motion $n$ expression becomes
\begin{equation}
    \label{eq:mean_motion}
    n^2 = 1 + \ell^2 + \frac{3A}{\left(1-e^2\right)^2}\left(1+\sqrt{1-e^2}\right) + \frac{2 M_b r_c}{\left({r_c}^2 + T^2\right)^{3/2}},
\end{equation}
with the term ${r_c}^2=1-\mu+\mu^2$ denotes the radial distance associated with the disk-like structure, as described in \citet{singh2014effects}.

\subsection{Potential \texorpdfstring{$W$}{W} approximation}

The model developed in this study is well-suited for application to systems such as asteroids or dwarf planets orbiting a star, where the mass ratio between the primary and secondary bodies is extremely small. This characteristic is exemplified within our own Solar System, considering the Sun as one of the primary bodies. For very small $\mu$, Eq. (\ref{eq:potential_W}) can be approximated with a series in the first order of $\mu$,
\begin{equation}
    \label{eq:W_approx}
    W \approx c_0 + c_1 \mu, 
\end{equation}
with
\begin{equation}
\begin{split}
    c_0=&\frac{1}{\sqrt{1-e^2}}\Bigg[\frac{1}{2}\left(x^2+y^2\right) + \frac{1}{\gamma_0}\Bigg(\frac{m_1 q \left(A+2 R^2\right)}{2R^3} \\
    &+ \frac{m_2}{2\ell} \log \left(\frac{\gamma_1+\gamma_2+2 \ell}{\gamma_1+\gamma_2-2 \ell}\right) + \frac{M_b}{\sqrt{R^2+T^2}}\Bigg)\Bigg],
\end{split}
\end{equation}
\begin{equation}
\begin{split}
    c_1=&\frac{1}{\sqrt{1-e^2}}\left(\frac{1}{\gamma_0}\right)\Bigg[\frac{m_1 q x \left(\frac{3A}{2}+\left(x^2+y^2\right)\right)}{R^5} + \frac{m_2 \left(\ell \left(\gamma_1-\gamma_2\right)+(x+1) \left(\gamma_1+\gamma_2\right)\right)}{\gamma_1 \gamma_2 \left(1-\ell^2+2x+x^2+y^2+\gamma_1 \gamma_2\right)} \\
    &+ \gamma_3\left(\frac{m_1 q \left(A+2 R^2\right)}{R^3} + \frac{2 M_b}{\sqrt{R^2+T^2}}\right.\left. + \frac{m_2}{\ell} \log \left(\frac{\gamma_1+\gamma_2+2 \ell}{\gamma_1+\gamma_2-2 \ell}\right)\right)\Bigg],
\end{split}
\end{equation}
with $\gamma_0$, $\gamma_1$, and $\gamma_2$ are described in Appendix \ref{app:def_gamma}.

It has to be noted that, in the context of the ERTBP, the eccentricity parameter $e$ refers explicitly to the mutual elliptical orbit of the two primaries (the larger and smaller interacting bodies) around their common center of mass. However, for systems with a very small mass ratio, the center of mass of the primaries lies extremely close to the center of the dominant primary. In this case, the motion of the smaller primary is effectively an elliptical orbit around the dominant central mass, while the dominant primary remains nearly stationary relative to the barycenter of the system.

\section{Location of the equilibrium points}
\label{sect:location}

In solving for equilibrium points, it is known that the points will be at rest in a rotating reference frame. This means that the velocities ($\dot{x}, \dot{y}$) and accelerations ($\ddot{x}, \ddot{y}$) at these points will be equal to zero. With this, the equations of motion can be simplified to obtain solutions of these equilibrium points under the condition that
\begin{align}
    \label{1st_derivative_V}
    &\frac{\partial{W}}{\partial{x}} = W_x = 0,  \\
    &\frac{\partial{W}}{\partial{y}} = W_y = 0. \notag
\end{align}
Thus, the first derivative of the potential is obtained with
\begin{equation}
\begin{split}
    \label{eq:pot_eq_zero_x}
    W_x = &\frac{1}{\sqrt{1-e^2}}\Bigg[x+\frac{1}{2\gamma_0}\Bigg(\Big(2\mu \gamma_3 + 1\Big) \Bigg(\frac{4 m_1 q x}{R^3} - \frac{3 m_1 q x \left(A+2 R^2\right)}{R^5} \\
    &- \frac{2 M_b x}{\left(T^2+R^2\right)^{3/2}} - \frac{4 m_2}{\left(\gamma_1+\gamma_2\right)^2-4\ell^2} \left(\frac{1+x-\ell}{\gamma_1}+\frac{1+x+\ell}{\gamma_2}\right)\Bigg) \\
    &+ 2\mu\Bigg(\frac{m_1 q \left(\frac{3 A}{2}+3 R^2\right)}{R^5} -\frac{5 m_1 q x^2 \left(3 A+2 R^2\right)}{2 R^7} + \frac{m_2 \gamma_4}{\gamma_1 \gamma_2 (\gamma_6)^2}\Bigg)\Bigg)\Bigg] = 0,
\end{split}
\end{equation}
\begin{equation}
\begin{split}
    \label{eq:pot_eq_zero_y}
    W_y = &\frac{1}{\sqrt{1-e^2}}\Bigg[y+\frac{1}{2\gamma_0}\Bigg(\Big(2\mu \gamma_3+1\Big) \Bigg(\frac{4 m_1 q y}{R^3} - \frac{3 m_1 q y \left(A+2 R^2\right)}{R^5} \\
    &- \frac{2 M_b y}{\left(T^2+R^2\right)^{3/2}} - \frac{2 m_2 y \left(\gamma_1+\gamma_2\right)}{\gamma_6}\Bigg) + 2\mu\Bigg(\frac{2 m_1 q x y}{R^5} \\
    &-\frac{5 m_1 q x y \left(3 A+2 R^2\right)}{2 R^7} + \frac{2 m_2 y \gamma_5}{\gamma_1 \gamma_2 (\gamma_6)^2}\Bigg)\Bigg)\Bigg] = 0.
\end{split}
\end{equation}

The collinear points align with the primaries, hence $y = 0$. Eq. (\ref{eq:pot_eq_zero_x}) becomes
\begin{equation}
\begin{split}
    \label{eq:Wx0}
    x+&\frac{1}{2\gamma_0}\Bigg(\Big(2\mu \gamma_3 + 1\Big) \Bigg(\frac{4 m_1 q x}{|x|^3} - \frac{3 m_1 q x \left(A+2 x^2\right)}{|x|^5} - \frac{2 M_b x}{\left(T^2+x^2\right)^{3/2}} \\
    &- \frac{2 m_2}{1-\ell^2+2x+x^2+|1+x-\ell||1+x+\ell|} \left(\frac{1+x-\ell}{|1+x-\ell|}+\frac{1+x+\ell}{|1+x+\ell|}\right)\Bigg) \\
    &+ 2\mu\Bigg(\frac{m_1 q \left(\frac{3 A}{2}+3 x^2\right)}{|x|^5} -\frac{5 m_1 q x^2 \left(3 A+2 x^2\right)}{2|x|^7} + \frac{m_2 \gamma_7}{\gamma_8}\Bigg)\Bigg) = 0.
\end{split}
\end{equation}
For non-collinear points, the additional condition is $y \neq 0$. Thus, the Eqs.(\ref{eq:pot_eq_zero_x}) and (\ref{eq:pot_eq_zero_y}) can be reformulated to take the following structure

\begin{equation}
\begin{split}
    \label{eq:tri_zero_x}
    &x \Bigg[1+\frac{1}{2\gamma_0} \Bigg( \Big(2\mu\gamma_3 + 1\Big)\Bigg(\frac{4m_1q}{R^3} -\frac{3m_1q\left(A+2R^2\right)}{R^5} -\frac{2M_b}{(T^2+R^2)^{3/2}} -\frac{2m_2(\gamma_1+\gamma_2)}{\gamma_6}\Bigg) \\
    &+ 2\mu \Bigg( \frac{2m_1qx}{R^5} -\frac{5m_1qx\left(3A+2R^2\right)}{2R^7}\Bigg)\Bigg)\Bigg] +\frac{1}{2\gamma_0}\Bigg(\Big(2\mu\gamma_3 + 1\Big) \\
    &\times\Bigg(\frac{2m_2\big((\gamma_1+\gamma_2)+\ell(\gamma_1-\gamma_2)\big)}{\gamma_6}\Bigg) + 2\mu\Bigg(\frac{m_1q\left(3A+2R^2\right)}{2R^5} + \frac{m_2 \gamma_9}{\gamma_1 \gamma_2 (\gamma_6)^2}\Bigg)\Bigg) = 0,
\end{split}
\end{equation}
\begin{equation}
\begin{split}
    \label{eq:tri_zero_y}
    &y\Bigg[1+\frac{1}{2\gamma_0}\Bigg(\Big(2\mu \gamma_3+1\Big) \Bigg(\frac{4 m_1 q}{R^3} - \frac{3 m_1 q \left(A+2 R^2\right)}{R^5} - \frac{2 M_b}{\left(T^2+R^2\right)^{3/2}} \\
    &- \frac{2 m_2 \left(\gamma_1+\gamma_2\right)}{\gamma_6}\Bigg) + 2\mu\Bigg(\frac{2 m_1 q x}{R^5} -\frac{5 m_1 q x \left(3 A+2 R^2\right)}{2 R^7} + \frac{2 m_2 \gamma_5}{\gamma_1 \gamma_2 (\gamma_6)^2}\Bigg)\Bigg)\Bigg] = 0.
\end{split}
\end{equation}
Since $y\neq0$, we get 
\begin{equation}
\begin{split}
    \label{eq:tri_zero_y_2}
    &1+\frac{1}{2\gamma_0}\Bigg(\Big(2\mu \gamma_3+1\Big) \Bigg(\frac{4 m_1 q}{R^3} - \frac{3 m_1 q \left(A+2 R^2\right)}{R^5} - \frac{2 M_b}{\left(T^2+R^2\right)^{3/2}} \\
    &- \frac{2 m_2 \left(\gamma_1+\gamma_2\right)}{\gamma_6}\Bigg) + 2\mu\Bigg(\frac{2 m_1 q x}{R^5} -\frac{5 m_1 q x \left(3 A+2 R^2\right)}{2 R^7} + \frac{2 m_2 \gamma_5}{\gamma_1 \gamma_2 (\gamma_6)^2}\Bigg)\Bigg)\Bigg] = 0.
\end{split}
\end{equation}
Eq. (\ref{eq:tri_zero_y_2}) can be substituted for Eq. (\ref{eq:tri_zero_x}). Hence,
\begin{equation}
\begin{split}
    \label{eq:tri_zero_simple}
    &\Big(2\mu\gamma_3 + 1\Big)\Bigg(\frac{2m_2\big((\gamma_1+\gamma_2)+\ell(\gamma_1-\gamma_2)\big)}{\gamma_6}\Bigg) \\
    &+ 2\mu\Bigg(\frac{m_1q\left(3A+2R^2\right)}{2R^5} + \frac{m_2 \gamma_9}{\gamma_1 \gamma_2 (\gamma_6)^2}\Bigg) = 0,
\end{split}
\end{equation}
with $\gamma_3$ to $\gamma_9$ are described in Appendix \ref{app:def_gamma}.

\begin{figure}[htbp]
    \centering
    \centerline{\includegraphics[width=1\textwidth]{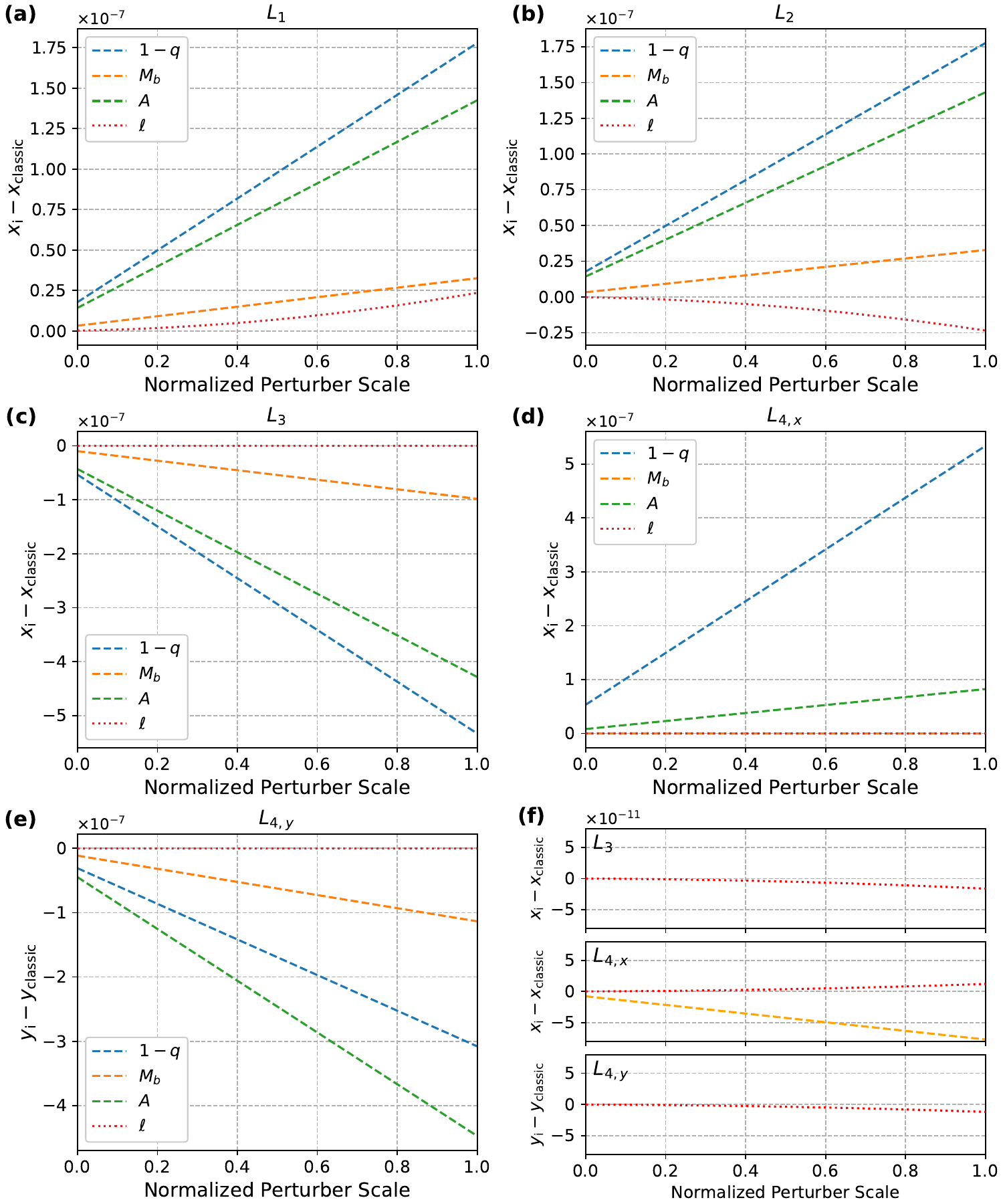}}
    \vspace{2mm}
    \caption{Effect of perturbation parameter variation on the position of equilibrium points. The original range scale for each perturbation parameter is $0$ to $1.6\times10^{-6}$ for $1-q$; $0$ to $2.6\times10^{-7}$ for $A$; $0$ to $7\times10^{-6}$ for $\ell$; and $0$ to $3\times10^{-7}$ for $M_b$. The ranges were then normalized to a scale of 0 to 1 for the $x$-axis. The $y$-axis is the positional difference between equilibrium points in the perturbed ERTBP and their location in the classical case. The last panel (f) is the zoomed-in version of $L_3$, $L_{4,x}$, and $L_{4,y}$ plots around $y=0$. For each variation of the perturbation parameter, the other parameters were set equal to $0$. We used $\mu=1\times10^{-9}$ and $e=0.2$.}
    \label{fig:effect_all}
\end{figure}

The locations of the collinear equilibrium points were calculated numerically by solving Eq. (\ref{eq:Wx0}) for $x$. Meanwhile, the locations of the non-collinear equilibrium points were calculated by solving Eqs. (\ref{eq:pot_eq_zero_x}) and (\ref{eq:tri_zero_y_2})  for $x$ and $y$. We performed this calculation using the Python SymPy package \citep{meurer2017sympy}. We found three collinear equilibrium points, $L_1$ located between the larger primary and the smaller primary ($\mu-1-\ell, \mu$) region, $L_2$ located in ($-\infty, \mu-1+\ell$) region, and $L_3$ located in ($\mu, \infty$) region. Meanwhile, two non-collinear equilibrium points were found, which are the non-collinear points $L_4$ and $L_5$. The effects of various perturbations are presented in Fig. \ref{fig:effect_all}. Here, the results are presented using a mass ratio of $1\times10^{-9}$ and an eccentricity of $0.2$. The figure only shows the non-collinear point $L_4$, but the trend of $L_5$ is identical, only mirrored symmetrically with respect to the $y$-axis.

Fig. \ref{fig:effect_all} plots the difference between equilibrium point positions on perturbed ERTBP and their classical case for different perturbation parameter variations in the context of a very small mass ratio. By incorporating the perturbation sources one by one, we have obtained some results. As the photogravitational parameter ($1-q$) increases, $L_1$ and $L_3$ move closer to the larger primary, while $L_2$ shifts toward the primaries, and $L_4$ shifts toward the larger primary and closer to the $x$-axis. When the oblateness parameter ($A$) increases, $L_1$ and $L_2$ shift toward the larger primary, whereas $L_3$ moves closer to the primaries, and $L_4$ moves toward the larger primary and closer to the $x$-axis. An increase in the length parameter of the finite straight segment ($\ell$) causes $L_1$ and $L_2$ to shift away from the smaller primary, while $L_3$ moves closer to the larger primary, and $L_4$ shifts closer to the larger primary and closer to the $x$-axis. As the total mass parameter of the disk-like structure ($M_b$) increases, $L_1$ shifts away and $L_2$ shifts closer to the smaller primary, $L_3$ moves closer to the primaries, and $L_4$ shifts away from the larger primary and closer to the $x$-axis.

In this work, a system was modeled using the restricted three-body problem framework and its equilibrium points and stability were calculated. The combination of parameters used to calculate the location of equilibrium points is divided into several cases, where:
\begin{enumerate}
    \label{enum:case}
    \item Case 1: includes all of the perturbation parameters we used.
    \item Case 2: the smaller primary is a point source ($\ell=0$).
    \item Case 3: there is no disk-like structure ($M_b = 0$).
    \item Case 4: the larger primary is spherical ($A=0$).
    \item Case 5: no radiation from the larger primary ($1-q=0$).
    \item Classic: no perturbation sources included ($e=1-q=A=M_b=\ell=0$).
\end{enumerate}
Case 1 includes all of the perturbation parameters we used, which are eccentricity ($e$), photogravitational effect ($q$), oblateness of the larger primary ($A$), the total mass of the disk-like structure ($M_b$), and the length of the finite straight segment of the smaller primary ($\ell$). From cases 2 to 5, one by one, the perturbation source was removed to examine the influence of the perturbation source on the displacement of the equilibrium point's position. Tables \ref{tab:loc-collinear} and \ref{tab:loc-triangular} present the shifts in the locations of the collinear and non-collinear equilibrium points caused by the respective perturbation cases. The shift is calculated as the difference between the perturbed value and the classical unperturbed value ($\Delta x = x_{\mathrm{perturbed}} - x_{\mathrm{classic}}$). The choice of parameters in Cases 1-5 is arbitrary to serve illustrative purposes only. The selected values are not excessively large so that the underlying approximations remain valid. The same parameter values are employed in all cases to enable a consistent evaluation of the effect of each perturbation parameter on the system.

\begin{sidewaystable}[htbp]
\centering
\small
\begin{minipage}[t]{0.87\textwidth}
\centering
\caption{Difference in location of the collinear equilibrium points between the perturbed ERTBP and the classical case ($\times 10^{-5}$) with $\mu=2\times10^{-9}$, $e=0.2$, $T=0.11$, and varying perturbation parameters}
\label{tab:loc-collinear}
{\renewcommand{\arraystretch}{1.2}
\begin{tabular*}{\linewidth}{@{\extracolsep{\fill}}cccccccc}
\toprule
Case & $1-q$ & $A$ & $M_b$ & $\ell$ & $L_1$ & $L_2$ & $L_3$ \\
\midrule
1 & $1\times 10^{-5}$     & $1\times 10^{-5}$  & $1\times 10^{-5}$ & $1\times 10^{-5}$ & $0.777870268176	$ & $0.761254859882$ & $-2.308767615333	$  \\
2 & $1\times 10^{-5}$     & $1\times 10^{-5}$  & $1\times 10^{-5}$ & $0$ & $0.774120572578$ & $0.765136786551$ & $-2.308764282388$  \\
3 & $1\times 10^{-5}$     & $1\times 10^{-5}$  & $0$ & $1\times 10^{-5}$ & $0.667477550764$ & $0.653424614705$ & $-1.981449450283$  \\
4 &$1\times 10^{-5}$     & $0$  & $1\times 10^{-5}$ & $1\times 10^{-5}$ & $0.224146536698$ & $0.216285214960$ & $-0.660701698629	$  \\
5 &$0$     & $1\times 10^{-5}$  & $1\times 10^{-5}$ & $1\times 10^{-5}$ & $0.664876975243$ & $0.652023674474$ & $-1.975446323743$  \\
\bottomrule
\end{tabular*}}
\end{minipage}

\par\bigskip  
\hspace{5mm}

\begin{minipage}[t]{0.87\textwidth}
\centering
\caption{Difference in location of the non-collinear equilibrium points between the perturbed ERTBP and the classical case ($\times 10^{-5}$) with $\mu=2\times10^{-9}$, $e=0.2$, $T=0.11$, and varying perturbation parameters}
\small
\label{tab:loc-triangular}
{\renewcommand{\arraystretch}{1.2}
\begin{tabular*}{\linewidth}{@{\extracolsep{\fill}}ccccccc}
\toprule
\multirow{2}{*}{Case} & \multirow{2}{*}{$1-q$} & \multirow{2}{*}{$A$} &
\multirow{2}{*}{$M_b$} & \multirow{2}{*}{$\ell$} & \multicolumn{2}{c}{$L_{4,5}$} \\
\cmidrule{6-7}
 & & & & & $x$ & $y$ \\
\midrule
1 & $1\times 10^{-5}$     & $1\times 10^{-5}$  & $1\times 10^{-5}$ & $1\times 10^{-5}$ & $-0.166677612912$ & $\pm2.762180008198$   \\
2 & $1\times 10^{-5}$     & $1\times 10^{-5}$  & $1\times 10^{-5}$ & $0$ & $-0.166676361424$ & $\pm2.762175437032$ \\
3 & $1\times 10^{-5}$     & $1\times 10^{-5}$  & $0$ & $1\times 10^{-5}$ & $-0.166675431690$ & $\pm2.384221166718$   \\
4 &$1\times 10^{-5}$     & $0$  & $1\times 10^{-5}$ & $1\times 10^{-5}$ & $0.333330457963$ & $\pm0.570464175942$   \\
5 &$0$     & $1\times 10^{-5}$  & $1\times 10^{-5}$ & $1\times 10^{-5}$ & $-0.499995002046$ & $\pm2.569738220903$   \\
\bottomrule
\end{tabular*}
\footnotetext{Note: $L_4$ is on the positive $y$-axis and $L_5$ is on the negative $y$-axis.\par}}
\end{minipage}
\end{sidewaystable}

From Table \ref{tab:loc-collinear}, the equilibrium points of $L_1$ and $L_2$ for the system show positive differences from the classical case. It means that in Cases 1-5, the location of $L_1$ (located between the larger and the smaller primary) shifts away from the smaller primary and the location of $L_2$ (on the far side of the smaller primary) moves closer to it compared to their respective classical locations. The equilibrium point of $L_3$ (on the far side of the larger primary) moves closer to the larger primary, as shown by negative differences compared to the classical case. The inclusion of all perturbation sources (Case 1) results in the largest difference at $L_1$ and $L_3$. 

By taking the differences of the equilibrium points' location between each case to the classical case and compared to the fully perturbed case (Case 1), removing the finite straight segment effect of the smaller primary ($\ell=0$, Case 2) produces virtually indistinguishable shifts (differences $<0.5\%$ at $L_{1,2}$ and $<0.0002\%$ at $L_3$). This result indicates that this factor contributes negligibly to the equilibrium displacements at this scale. 

When the disk-like structure or radiation pressure from the larger primary is omitted (Case 3 and Case 5), the magnitudes at all collinear equilibrium points drop to about a $15\%$ reduction relative to the Case 1 shift. It shows that the disk and the radiation pressure moderately influence the collinear equilibrium points. Radiation pressure acts as an effective reduction of the larger primary’s gravity. When radiation is included, the weakened attraction must be compensated for by moving the equilibrium points closer to the larger primary. Suppressing the oblateness of the larger primary (Case 4) produces the largest reduction ($\sim\!72\%$) relative to Case 1 deviation from the classical case for all collinear equilibrium points. This establishes oblateness $A$ as the dominant control parameter for collinear point displacements when all perturbation parameters are set equal. The effect of turning off oblateness for $L_{1,2,3}$ arises because the oblate primary adds a correction to the potential that enhances the in-plane radial gravity relative to a sphere. 

From Table \ref{tab:loc-triangular}, the negative difference in the $x$-axis of non-collinear points in the system relative to the classical case shows that the location of the non-collinear point moves away from the larger primary in Cases 1, 2, 3, and 5. For the case where the larger primary is spherical (Case 4), the location moves closer toward the larger primary. Meanwhile, the results on the $y$-axis show that the location moves closer to the abscissa in all cases. 

Across Cases 1–3, the difference in position in the $x$-axis for the non-collinear equilibrium points is around $-1.67\times10^{-6}$ from the classical case. Zeroing the finite segment length ($\ell=0$, Case 2) or removing the disk mass ($M_b=0$, Case 3) changes the shift by about $0.001\%$ from the Case 1 shift relative to the classical case. Thus, those two perturbations hardly influence the $x$-shift. By contrast, Cases 4 and 5 show that oblateness ($A$) and radiation pressure ($1-q$) have greater influence, acting in opposite directions. Physically, the oblateness of the larger primary shifts the non-collinear points to the left (more negative $x$). In comparison, radiation pressure effectively weakens the gravity of the larger primary and pushes the non-collinear points to the right. The near-cancellation of these two effects explains the modest offset observed when all perturbations are present, and why removing either one produces a more significant and opposite shift.

For the $y$-axis, inclusion of all perturbation sources (Case 1) creates the most significant margin with respect to the classical case. Setting the small primary to a point source (Case 2) perturbs the $y$-axis of non-collinear points only at $0.0002\%$ level compared to Case 1. It demonstrates that the finite segment length is dynamically negligible for the non-collinear height. Removing the disk-like component (Case 3) results in a substantial downward offset of approximately $13\%$ compared to Case 1. The dominant control on the $y$-shift in non-collinear points is the oblateness of the larger primary. When oblateness is suppressed (Case 4), the vertical displacement collapses approximately $79\%$ compared to the Case 1 shift, revealing that the oblateness effect is chiefly responsible for the pronounced depression of the non-collinear equilibrium points. Radiation pressure exerts a moderate influence, which shifts the $y$-axis of the non-collinear points about $7\%$ relative to the Case 1 shift.

Numerical analysis of the perturbed system, as summarized in Tables \ref{tab:loc-collinear} and \ref{tab:loc-triangular}, demonstrated that inclusion the parameters $q$, $A$, $\ell$, and $M_b$ produced measurable shifts in equilibrium point positions relative to the classical ERTBP. Specifically, Table \ref{tab:loc-collinear} highlights that the locations of the collinear points are sensitive to the varying perturbation parameters, showing distinct deviations compared to the classical case. Similarly, Table \ref{tab:loc-triangular} indicates that the coordinates of the non-collinear points $L_{4,5}$ undergo significant displacement when subjected to the combined effects of the perturbed parameters.

\section{Stability of the equilibrium points}
\label{sect:stability}

Two things may happen if an object is disturbed at an equilibrium point. If the motion resulting from the motion of the particle moves away from its equilibrium point, the equilibrium is unstable. If the particle only oscillates in the region around the equilibrium point, the equilibrium is stable \citep{roy2005orbital}. A slight displacement from these points is considered to evaluate the stability of the equilibrium points. Hence we define: $x = x_0 + \delta x$ and $y = y_0 + \delta y$ with $\delta x$ and $\delta y$ being infinitesimal. The equations of motion for these small displacements are derived from Eq. (\ref{eq:motion_eq}) as follows
\begin{equation}
    \begin{aligned}
    \label{eq:delta_x_delta_y}
    \delta\ddot{x} - 2n\delta\dot{y} &= W_{xx}\delta x + W_{xy} \delta y, \\
    \delta\ddot{y} + 2n\delta\dot{x} &= W_{xy}\delta x + W_{yy} \delta y. 
    \end{aligned}
\end{equation}
Eq. (\ref{eq:delta_x_delta_y}) has general solutions in the form of $\delta x = \delta x_0\, e^{\lambda t}$ and $\delta y = \delta y_0\, e^{\lambda t}$. Therefore, the characteristic equation corresponding to the equations is
\begin{equation} 
    \label{eq:characteristic}
    \lambda^4 + (4n^2 - W^0_{xx} - W^0_{yy})\lambda^2 + W^0_{xx}W^0_{yy} - (W^0_{xy})^2 = 0.
\end{equation}
The second partial derivatives of $W$ are denoted using subscripts. The superscript zero indicates that these derivatives were evaluated at the equilibrium point ($x_0, y_0$), where
\begin{equation}
    \begin{split}
    \label{eq:Wxx}
    W_{xx}=&\frac{1}{\sqrt{1-e^2}}\Bigg[1+\frac{1}{2\gamma_0} \Bigg(\big(2\mu \gamma_3+1\big)\Bigg(\frac{4 m_1 q}{R^3} -\frac{3 m_1 q \left(A+2 R^2\right)}{R^5} -\frac{24 m_1 q x^2}{R^5} \\
    &+\frac{15 m_1 q x^2 \left(A+2 R^2\right)}{R^7} -\frac{2 M_b}{\left(T^2+R^2\right)^{3/2}} +\frac{6 M_b x^2}{\left(T^2+R^2\right)^{5/2}} \\
    &+\frac{8 m_2 (\gamma_1+\gamma_2) \left(\frac{x+1-\ell}{\gamma_1}+\frac{x+1+\ell}{\gamma_2}\right)^2}{\left((\gamma_1+\gamma_2)^2-4 \ell^2\right)^2} -\frac{4 m_2 \left(\frac{1}{\gamma_1}+\frac{1}{\gamma_2}-\frac{(x+1-\ell)^2}{\gamma_1^3}-\frac{(x+1+\ell)^2}{\gamma_2^3}\right)}{(\gamma_1+\gamma_2)^2-4 \ell^2}\Bigg) \\
    &+2\mu \Bigg(\frac{2 m_1 q (x+1)}{R^5} -\frac{10 m_1 q x^2 (x+1)}{R^7} -\frac{15 m_1 q x \left(3 A+2 R^2\right)}{2 R^7} \\
    &+\frac{35 m_1 q x^3 \left(3 A+2 R^2\right)}{2 R^9} -\frac{\gamma_{10} m_2}{\gamma_1 \gamma_2 \gamma_6^2} +\frac{2 \gamma_4 m_2 \Big(\frac{\gamma_2 (x+1-\ell)}{\gamma_1}+\frac{\gamma_1 (x+1+\ell)}{\gamma_2}+2 x+2\Big)}{\gamma_6^3} \\
    &+\frac{3 \gamma_4 m_2 (x+1-\ell)}{\gamma_1^3 \gamma_2 \gamma_6^2} +\frac{3 \gamma_4 m_2 (x+1+\ell)}{\gamma_1 \gamma_2^3 \gamma_6^2} \Bigg)\Bigg)\Bigg],
    \end{split}
\end{equation}
\begin{equation}
    \begin{split}
    \label{eq:Wyy}
    W_{yy}=&\frac{1}{\sqrt{1-e^2}} \Bigg[1+\frac{1}{2\gamma_0} \Bigg(\big(2\mu \gamma_3+1\big)\Bigg(\frac{4 m_1 q}{R^3} -\frac{3 m_1 q \left(A+2 R^2\right)}{R^5} -\frac{24 m_1 q y^2}{R^5} \\
    &+\frac{15 m_1 q y^2 \left(A+2 R^2\right)}{R^7} -\frac{2 M_b}{\left(T^2+R^2\right)^{3/2}} +\frac{6 M_b y^2}{\left(T^2+R^2\right)^{5/2}} +\frac{2 m_2 y^2 (\gamma_1+\gamma_2)^3}{\gamma_6^2} \\
    &+\frac{2 m_2 y^2 (\gamma_1+\gamma_2)}{\gamma_1^2 \gamma_6} +\frac{2 m_2 y^2 (\gamma_1+\gamma_2)}{\gamma_2^2 \gamma_6} -\frac{2 m_2 (\gamma_1+\gamma_2) \left(\gamma_1 \gamma_2+y^2\right)}{\gamma_1 \gamma_2 \gamma_6} \Bigg) \\
    &+2\mu \Bigg(\frac{2 m_1 q x}{R^5} -\frac{20 m_1 q x y^2}{R^7} -\frac{5 m_1 q x \left(3 A+2 R^2\right)}{2 R^7} +\frac{35 m_1 q x y^2 \left(3 A+2 R^2\right)}{2 R^9} \\
    &-\frac{2 \gamma_{11} m_2 y^2}{\gamma_1 \gamma_2 \gamma_6^2} +\frac{4 \gamma_5 m_2 y^2 (\gamma_1+\gamma_2)^2}{\gamma_1 \gamma_2 \gamma_6^3} -\frac{2 \gamma_5 m_2}{\gamma_1 \gamma_2 \gamma_6^2}+\frac{6 \gamma_5 m_2 y^2}{\gamma_1^3 \gamma_2 \gamma_6^2}+\frac{6 \gamma_5 m_2 y^2}{\gamma_1 \gamma_2^3 \gamma_6^2}  \Bigg)\Bigg)\Bigg],
    \end{split}
\end{equation}	
\begin{equation}
    \begin{split}
    \label{eq:Wxy}
    W_{xy}=&\frac{1}{\sqrt{1-e^2}} \Bigg(\frac{y}{2\gamma_0}\Bigg) \Bigg[\big(2\mu \gamma_3+1\big)\Bigg(-\frac{24 m_1 q x}{R^5} +\frac{15 m_1 q x \left(A+2 R^2\right)}{R^7} \\
    &+\frac{6 M_b x}{\left(T^2+R^2\right)^{5/2}} +\frac{8 m_2 (\gamma_1+\gamma_2)^2 \left(\frac{x+1-\ell}{\gamma_1}+\frac{x+1+\ell}{\gamma_2}\right)}{\gamma_1 \gamma_2 \left((\gamma_1+\gamma_2)^2-4 \ell^2\right)^2} +\frac{4 m_2 \left(\frac{x+1-\ell}{\gamma_1^3}+\frac{x+1+\ell}{\gamma_2^3}\right)}{(\gamma_1+\gamma_2)^2-4 \ell^2} \Bigg) \\
    &+2\mu\Bigg(\frac{2 m_1 q}{R^5} -\frac{10 m_1 q x (x+1)}{R^7} -\frac{5 m_1 q \left(3 A+2 R^2\right)}{2 R^7} +\frac{35 m_1 q x^2 \left(3 A+2 R^2\right)}{2 R^9} \\
    &-\frac{\gamma_{12} m_2}{\gamma_1 \gamma_2 \gamma_6^2} +\frac{2 \gamma_4 m_2 (\gamma_1+\gamma_2)^2}{\gamma_1 \gamma_2 \gamma_6^3} +\frac{3 \gamma_4 m_2}{\gamma_1^3 \gamma_2 \gamma_6^2}+\frac{3 \gamma_4 m_2}{\gamma_1 \gamma_2^3 \gamma_6^2} \Bigg)\Bigg],
    \end{split}
\end{equation}
with $\gamma_{10}$, $\gamma_{11}$, and $\gamma_{12}$ are described in Appendix \ref{app:def_gamma}.

The characteristic equation has roots of 
\begin{equation}
    \label{eq:char_roots}
    \lambda_{1,2,3,4} = \pm \sqrt{\frac{-b \pm \sqrt{b^2 - 4ac}}{2a}}\,,
\end{equation}
where $a = 1$, $b = 4n^2 - W^0_{xx} - W^0_{yy}$, and $c = W^0_{xx}W^0_{yy} - (W^0_{xy})^2$.
When all roots are purely imaginary and appear in conjugate pairs, this would result in a ``center $\times$ center" dynamical behavior, indicating a stable system condition. If the system has two purely imaginary roots along with a pair of real roots, it results in the presence of a positive eigenvalue, which renders the system unstable. This instability causes trajectories to diverge in certain directions while exhibiting oscillatory behavior in others, defining the ``saddle $\times$ center'' dynamic \citep{mahato2022effect}.

\begin{sidewaystable}[htbp]
\centering
\setlength{\tabcolsep}{2.5pt} 
\renewcommand{\arraystretch}{1.3}
\caption{Characteristic roots of the equilibrium points with $\mu=2\times10^{-9}$, $A=1-q=\ell=M_b=1\times10^{-5}$, $T=0.11$, and varying eccentricities}
\label{tab:characteristic_roots}
\begin{tabular*}{\linewidth}{@{\extracolsep{\fill}} c cc cc cc cc}
\toprule
\multirow{2}{*}{$e$} & \multicolumn{2}{c}{$L_1$} & \multicolumn{2}{c}{$L_2$} & \multicolumn{2}{c}{$L_3$} & \multicolumn{2}{c}{$L_{4,5}$} \\
\cmidrule(lr){2-3} \cmidrule(lr){4-5} \cmidrule(lr){6-7} \cmidrule(l){8-9}
& $\lambda_{1,2}$ & $\lambda_{3,4}$ & $\lambda_{1,2}$ & $\lambda_{3,4}$ & $\lambda_{1,2}$ & $\lambda_{3,4}$ & $\lambda_{1,2}$ & $\lambda_{3,4}$ \\
\midrule
0.2     & $\pm2.513957$ & $\pm2.066684i$ & $\pm2.573782$ & $\pm2.103423i$ & $\pm0.000076$ & $\pm0.968733i$ & $\pm0.000122i$ & $\pm0.968733i$ \\
0.5     & $\pm2.716129$ & $\pm2.139969i$ & $\pm2.812533$ & $\pm2.201976i$ & $\pm0.000114$ & $\pm0.732353i$ & $\pm0.000183i$ & $\pm0.732353i$ \\
0.6613  & $\pm2.950673$ & $\pm2.226273i$ & $\pm3.119794$ & $\pm2.340098i$ & $\pm0.002622$ & $\pm0.036841i$ & $\pm0.004244i$ & $\pm0.036502i$ \\
0.6616  & $\pm2.951241$ & $\pm2.226480i$ & $\pm3.120596$ & $\pm2.340475i$ & $\pm0.011451$ & $\pm0.008439	i$ & $\pm0.009615\pm0.007906i$ & $\pm0.009615\mp0.007906i$ \\
0.8     & $\pm3.285684$ & $\pm2.339020i$ & $\pm3.696294$ & $\pm2.631343i$ & $\pm0.999244$ & $\pm0.000121i$ & $\pm0.99924$ & $\pm0.000194$ \\
\bottomrule
\end{tabular*}
\footnotetext{Note: $i=\sqrt{-1}\,.$\par}
\end{sidewaystable}

\begin{figure}
    \centering
    \includegraphics[width=0.99\textwidth,keepaspectratio]{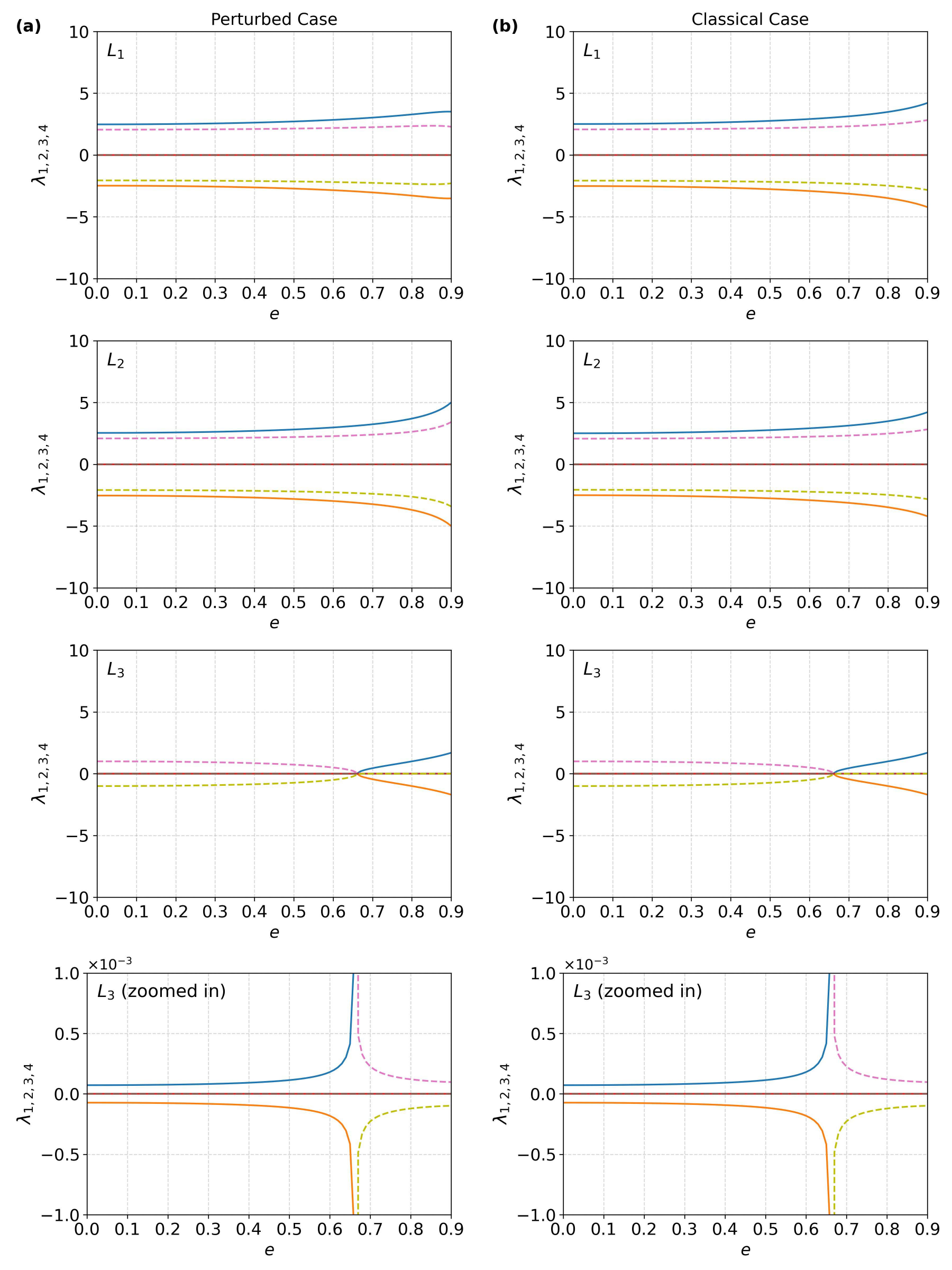}
    \caption{Plot of eccentricity versus characteristic roots ($\lambda_{1,2,3,4}$) for collinear equilibrium points ($L_1$, $L_2$, and $L_3$) for perturbed (a) and classical (b) cases. The details of the parameters used in each subfigure are as follows: (a) $\mu=2\times 10^{-9}$, $A=1-q=\ell=M_b=1\times10^{-5}$, and $T=0.11$; (b) $\mu=2\times10^{-9}$ and $A=1-q=\ell=M_b=T=0$. The bottommost panels show the zoomed-in version of $L_3$ eigenplot. The real and imaginary parts of the characteristic roots are shown by solid and dashed lines, respectively.}
    \label{fig:eigen_col_plot}
\end{figure}

\begin{figure}
    \centering
    \includegraphics[width=1\textwidth,keepaspectratio]{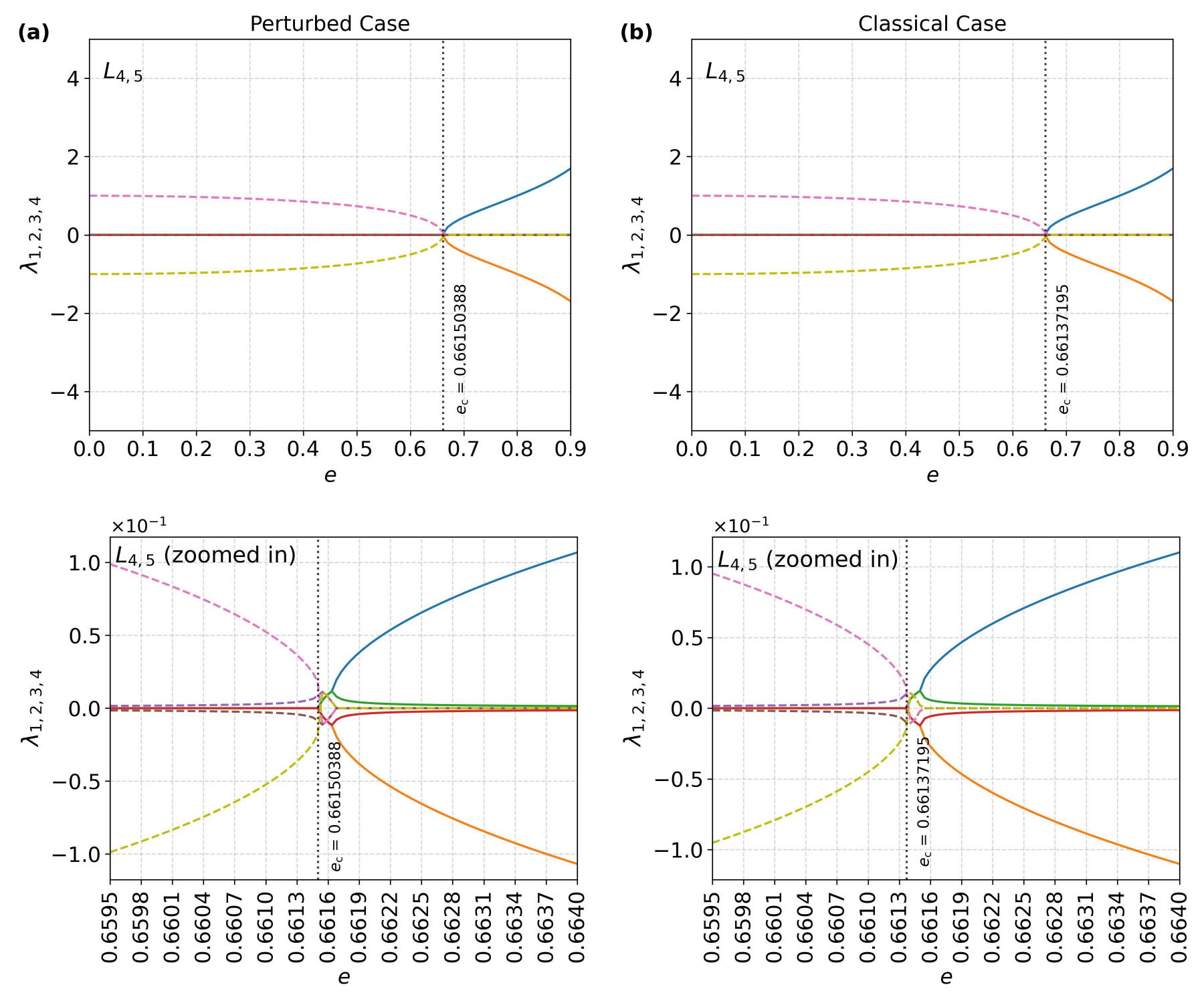}
    \caption{Plot of eccentricity versus characteristic roots ($\lambda_{1,2,3,4}$) for non-collinear equilibrium points ($L_4$ and $L_5$) for perturbed (a) and classical (b) cases. Zoomed-in versions are shown in the bottom subfigures. The details of the parameters used in each subfigure are as follows: (a) $\mu=2\times 10^{-9}$, $A=1-q=\ell=M_b=1\times10^{-5}$, and $T=0.11$; (b) $\mu=2\times10^{-9}$ and $A=1-q=\ell=M_b=T=0$. The real and imaginary parts of the characteristic roots are shown by solid and dashed lines, respectively.}
    \label{fig:eigen_tri_plot}
\end{figure}

Collinear equilibrium points are located along the abscissa, where the condition $W_{xy} = 0$ is satisfied. To investigate their stability, the abscissa is partitioned into three distinct regions: $L_1 = (-\infty, \mu-1-\ell)$, $L_2 = (\mu-1-\ell, \mu)$, and $L_3 = (\mu, \infty)$. The values of the characteristic roots $\lambda$ within each region were numerically evaluated to determine their values. Similarly, for the non-collinear equilibrium points ($L_4$ and $L_5$), located off the abscissa, the mixed derivative term $W_{xy}^0$ generally does not vanish. Investigating their stability requires evaluating the second-order partial derivatives ($W_{xx}^0, W_{yy}^0,$, and $W_{xy}^0$) at specific coordinate values corresponding to $L_4$ and $L_5$, and substituting these into the coefficients $b$ and $c$ to determine the nature of the roots numerically.

The stability analysis begins by incorporating the perturbation parameters ($1-q$, $A$, $\ell$, and $M_b$) from Case 1 in Tables \ref{tab:loc-collinear} and \ref{tab:loc-triangular}, as well as for the classical case. The numerical results are summarized in Table \ref{tab:characteristic_roots}, and visualized in Figs. \ref{fig:eigen_col_plot} and \ref{fig:eigen_tri_plot}, which compare between the fully perturbed and the classical cases. Table \ref{tab:characteristic_roots} lists the characteristic roots ($\lambda_{1,2,3,4}$) for selected eccentricity values, providing numerical evidence for the stability transitions for the fully perturbed case. For collinear equilibrium points (Fig. \ref{fig:eigen_col_plot}), it was revealed that a pair of real nonzero eigenvalues appear with opposite signs, and a pair of purely imaginary eigenvalues also occur with opposite signs for eccentricity values within the range of $0\leq e < 1$. This is confirmed by Table \ref{tab:characteristic_roots}, which shows real roots for $L_1$ and $L_2$ at all tabulated eccentricities. Although the values around $L_3$ may appear to approach zero until $e < 0.7$, they retain nonzero magnitudes. For instance, at $e=0.2$, Table \ref{tab:characteristic_roots} shows small but finite real parts ($\lambda_{1,2}=\pm0.000076$). The bottommost panel of Fig. \ref{fig:eigen_col_plot} provides a clearer illustration by presenting a zoomed-in visualization of the $L_3$ eigenplot, confirming these results. Consequently, the collinear equilibrium points exhibit a ``saddle $\times$ center" behavior. This observation suggests that all collinear equilibrium points in both the perturbed and classical cases are intrinsically unstable under these conditions.

The stability of the non-collinear equilibrium points was examined using Eqs. (\ref{eq:characteristic})-(\ref{eq:Wxy}) and the data in Table \ref{tab:characteristic_roots}. We observed that two pairs of purely imaginary eigenvalues appear with opposite signs for a certain range of eccentricity ($0 \leq e \leq e_c$), as illustrated in Fig. \ref{fig:eigen_tri_plot} for the perturbed and classical cases. Table \ref{tab:characteristic_roots} highlights this stable region, showing purely imaginary roots for $L_{4,5}$, for instance at $e=0.6613$, $\lambda_{1,2}=\pm0.004244i$ and $\lambda_{3,4}=\pm0.036502i$. However, beyond the critical point $e_c$, these eigenvalues transition to a pair of real nonzero eigenvalues and a pair of imaginary eigenvalues. This bifurcation is evident in both Table \ref{tab:characteristic_roots} and Fig. \ref{fig:eigen_tri_plot}, where the characteristic roots become complex with real part near $e \approx 0.6616$ (e.g., $\lambda_{1,2} = \pm0.009615\pm0.007906i$ for the fully perturbed case). Thus, the non-collinear equilibrium points exhibit ``center $\times$ center'' and ``saddle $\times$ center'' behaviors with respect to eccentricity, leading to distinct stability characteristics in different regions. 

Specifically, for the region where $0 \leq e \leq e_c$, the non-collinear equilibrium points remain stable, whereas in the region where $e_c < e < 1$, these points become unstable. For the perturbed case (using $\mu=2\times10^{-9}$ and $A=1-q=\ell=M_b=1\times10^{-5}$), the critical eccentricity is $0.66150388$ and for the classical case (using $\mu=2\times10^{-9}$ and $A=1-q=\ell=M_b=0$), it is $0.66137195$. This bifurcation in stability at the critical eccentricity $e_c$ underscores the significant influence of eccentricity on the dynamical behavior of the system. We find the critical eccentricities for the parameter sets used in our examples to be very close in the perturbed and classical systems using the same mass ratio. The change in $e_c$ confirms that the perturbation parameters directly affect the stability limits.

\section{Conclusions}
\label{sect:conclusion}

In this study, we developed a dynamical model for systems characterized by an extremely small mass ratio between the primary bodies. We analyzed the motion of an infinitesimal mass influenced by two primaries under multiple perturbations, specifically considering an oblate and radiating larger primary, an elongated smaller primary approximated by a finite straight segment, and the gravitational influence of a surrounding disk-like structure. Our investigation of this modified ERTBP revealed five equilibrium points: three collinear ($L_1$, $L_2$, $L_3$) and two non-collinear ($L_4$, $L_5$). As shown in Tables \ref{tab:loc-collinear} and \ref{tab:loc-triangular}, numerical analysis of the perturbed system demonstrated that inclusion of orbital eccentricity and the parameters $q$, $A$, $\ell$, and $M_b$ produced measurable shifts in equilibrium point positions relative to the classical ERTBP. The magnitudes of these positional deviations varied with the model parameters.

Our stability analysis, detailed in Table \ref{tab:characteristic_roots} and visualised in Figs. \ref{fig:eigen_col_plot} and \ref{fig:eigen_tri_plot}, revealed that all three collinear equilibrium points remain linearly unstable across the broad range of eccentricities we examined. In contrast, the non-collinear points exhibit conditional stability depending on eccentricity. Furthermore, we identified critical eccentricity values ($e_c$) of  the perturbed and the classical systems, within which the non-collinear equilibrium points remain stable. The chosen perturbation parameters directly influence these stability limits.

To emphasize the role of our work in advancing the understanding of the ERTBP, it is essential to highlight how our findings contribute to the broader context of this field. Our study provides a comprehensive analysis of the effect of perturbation on the shift of the equilibrium point position with respect to eccentricity $e$ and the mass ratio $\mu$. Our study also provides a detailed analysis of the stability and dynamical behaviors of collinear and non-collinear equilibrium points across varying eccentricities, particularly identifying the critical threshold $e_c$ where stability transitions occur.

Our results show that the radiative forces and oblateness of the larger primary, the elongated-body geometry of the smaller primary, and the surrounding disk potentials have a significant impact on the dynamics of the third body, including its equilibrium points location and stability. The modified ERTBP model can serve as a good initial approximation for studying the dynamics of the restricted three-body problem in Solar System cases, such as Sun-asteroid or Sun-dwarf-planet systems, which are characterized by very small mass ratios.


\backmatter

\bmhead{Acknowledgements}

MBS is supported by the Indonesian Endowment Fund (LPDP) through a scholarship for master's studies. HSR is supported by Hibah Publikasi Pascasarjana FMIPA UI No.~PKS-042/UN2.F3.D/PPM.00.02/2024. INH is partially supported by RIIM LPDP-BRIN. We thank the reviewers for their time and insightful comments, which significantly improved the manuscript.

\section*{Declarations}


\begin{itemize}
\item Conflict of interest:
The authors have no competing interests to declare that are relevant to the content of this article.
\item Data availability: 
No datasets were generated or analyzed during the current study.
\item Author contribution:
HSR and INH conceived the research idea. HSR provided with analytical framework for calculations. INH provided the numerical help for the calculations. MBS calculated analytically and numerically the observables obtained. LBP helped with the numerical computations and physical insights. All authors reviewed and approved the manuscript.
\end{itemize}


\clearpage





\begin{appendices}

\section{Definition of \texorpdfstring{$\gamma$}{gamma} variables}\label{app:def_gamma}

These $\gamma$ variables from Eqs. (\ref{eq:g0})-(\ref{eq:g12}) are used to simplify Eqs. (\ref{eq:W_approx})-(\ref{eq:Wxy}).

\begin{align}
\boldsymbol{\gamma_0} &= 1+\ell^2+\frac{3A\left(1+\sqrt{1-e^2}\right)}{\left(1-e^2\right)^2}+\frac{2 M_b}{\left(1+T^2\right)^{3/2}} \label{eq:g0} \\[7pt]
\boldsymbol{\gamma_1} &= \sqrt{(1+x-\ell)^2+y^2} \\[7pt]
\boldsymbol{\gamma_2} &= \sqrt{(1+x+\ell)^2+y^2} \\[7pt]
\boldsymbol{\gamma_3} &= \frac{2 M_b \left(T^2-2\right)}{\left(T^2+1\right) \left(4 M_b+ \left(3 A+2 \ell^2+2\right)\left(T^2+1\right)^{3/2}\right)} \\[7pt]
\boldsymbol{\gamma_4} &= \begin{aligned}[t]
 &-4 \ell^5 (x+1) \big(\gamma_1-\gamma_2\big) -4 \ell^4 (x+1)^2 \left(\gamma_1+\gamma_2\right) \\
    &+ 2 \ell^3 (x+1) \left(4 (x+1)^2+y^2\right) \big(\gamma_1-\gamma_2\big) \\
    &+ 2 \ell^2 \Big(3 (x+1)^2 y^2 +4 (x+1)^4+y^4\Big) \big(\gamma_1+\gamma_2\big) \\
    &- 2 \ell (x+1) \Big((x+1)^2 y^2 +2 (x+1)^4-y^4\Big) \big(\gamma_1-\gamma_2\big) \\
    &- 2 \Big(2 (x+1)^2-y^2\Big)\Big((x+1)^2+y^2\Big)^2 \big(\gamma_1+\gamma_2\big)
\end{aligned}\\[7pt]
\boldsymbol{\gamma_{5}} &= \begin{aligned}[t]
&\ell^5 \left(\gamma_2-\gamma_1\right) - \ell^4 (x+1) \left(\gamma_1+\gamma_2\right) - \ell^3 \left(2 (x+1)^2+y^2\right) \left(\gamma_1-\gamma_2\right) \\
&- 2 \ell^2 (x+1)^3 \left(\gamma_1+\gamma_2\right) + 3 \ell (x+1)^2 \left((x+1)^2+y^2\right) \left(\gamma_1-\gamma_2\right) \\
&+ 3 (x+1) \left((x+1)^2+y^2\right)^2 \left(\gamma_1+\gamma_2\right)
\end{aligned} \\[7pt] 
\boldsymbol{\gamma_{6}} &= \gamma_1 \gamma_2 \left(1-\ell^2+2x+x^2+y^2+\gamma_1 \gamma_2\right) \\[7pt]
\boldsymbol{\gamma_{7}} &= \begin{aligned}[t]
&4(x+1)(1-\ell^2+2x+x^2)^2 (x|1+x-\ell| + \ell|1+x-\ell| + |1+x-\ell| \\
&+ x|1+x+\ell| - \ell|1+x+\ell|+|1+x+\ell|)
\end{aligned} \\[7pt] 
\boldsymbol{\gamma_{8}} &= (|1+x-\ell|)^3 (|1+x+\ell|)^3(1-\ell^2+2x+x^2 +|1+x-\ell||1+x+\ell|)^2 \\[7pt] 
\boldsymbol{\gamma_{9}} &= \begin{aligned}[t]
&2\left(\gamma_1-\gamma_2\right) \Big(-\ell^5 (x+2)  +\ell^3 \left(2 x \big(3 x^2+8x+y^2+7\right) +y^2+4\big) \\
&\qquad\qquad\quad\ +\ell^2 \big(3 (x+1)^2 y^2 +4 (x+1)^4+y^4\big) \\
&\qquad\qquad\quad\ - \ell(x+1) \left(5x^2+7x-y^2+2\right)\left((x+1)^2+y^2\right)\Big) \\
&+2\left(\gamma_1+\gamma_2\right) \Big(-\left(\ell^4 (x+1) (x+2)\right) +2 \ell^2 x (x+1)^3 \\
&\qquad\qquad\qquad\ -\left(5x^2+7x-y^2+2\right)\left((x+1)^2+y^2\right)^2\Big) \end{aligned} \\[7pt]
\boldsymbol{\gamma_{10}} &= \begin{aligned}[t]
	&-4 \ell^5 (x+1) \bigg(\frac{x+1-\ell}{\gamma_1}-\frac{x+1+\ell}{\gamma_2}\bigg) -4 \ell^5 (\gamma_1-\gamma_2) \\
	&-4 \ell^4 (x+1)^2 \bigg(\frac{x+1-\ell}{\gamma_1} +\frac{x+1+\ell}{\gamma_2}\bigg) -8 \ell^4 (x+1) (\gamma_1+\gamma_2) \\
	&+16 \ell^3 (x+1)^2 (\gamma_1-\gamma_2) +2 \ell^3 (x+1) \left(4 (x+1)^2+y^2\right) \\
	&\qquad \times \bigg(\frac{x+1-\ell}{\gamma_1} -\frac{x+1+\ell}{\gamma_2}\bigg) \\
	&+2 \ell^3 (\gamma_1-\gamma_2) \left(4 (x+1)^2+y^2\right) \\
	&+2 \ell^2 (\gamma_1+\gamma_2) \left(6 (x+1) y^2+16 (x+1)^3\right) \\
	&+2 \ell^2 \left(3 (x+1)^2 y^2+4 (x+1)^4+y^4\right) \bigg(\frac{x+1-\ell}{\gamma_1}+\frac{x+1+\ell}{\gamma_2}\bigg) \\
	&-2 \ell (x+1) (\gamma_1-\gamma_2) \left(2 (x+1) y^2+8 (x+1)^3\right) \\
	&-2 \left(2 (x+1)^2-y^2\right) \left((x+1)^2+y^2\right)^2 \left(\frac{x+1-\ell}{\gamma_1}+\frac{x+1+\ell}{\gamma_2}\right) \\
	&-2 \ell (x+1) \left((x+1)^2 y^2+2 (x+1)^4-y^4\right) \left(\frac{x+1-\ell}{\gamma_1}-\frac{x+1+\ell}{\gamma_2}\right) \\
	&-2 \ell (\gamma_1-\gamma_2) \left((x+1)^2 y^2+2 (x+1)^4-y^4\right) \\
	&-8 (x+1) (\gamma_1+\gamma_2) \left((x+1)^2+y^2\right)^2 \\
	&-8 (x+1) (\gamma_1+\gamma_2) \left(2 (x+1)^2-y^2\right) \left((x+1)^2+y^2\right) \end{aligned} \\[15pt]
\boldsymbol{\gamma_{11}} &= \begin{aligned}[t]
	&\ell^5 \left(\frac{1}{\gamma_2}-\frac{1}{\gamma_1}\right)+\ell^4 (-x-1) \left(\frac{1}{\gamma_1}+\frac{1}{\gamma_2}\right) \\
	&+\ell^3 \bigg(\left(\frac{1}{\gamma_1}-\frac{1}{\gamma_2}\right) \left(-2 (x+1)^2-y^2\right) -2 (\gamma_1-\gamma_2)\bigg) \\
	&-2 \ell^2 (x+1)^3 \left(\frac{1}{\gamma_1}+\frac{1}{\gamma_2}\right) \\
	&+\ell \Big(6 (x+1)^2 (\gamma_1-\gamma_2) +3 (x+1)^2 \left(\frac{1}{\gamma_1}-\frac{1}{\gamma_2}\right) \left((x+1)^2+y^2\right)\Big) \\
	&+3 (x+1) \left(\frac{1}{\gamma_1}+\frac{1}{\gamma_2}\right) \left((x+1)^2+y^2\right)^2 \\
	&+12 (x+1) (\gamma_1+\gamma_2) \left((x+1)^2+y^2\right) \end{aligned} \\[15pt]
\boldsymbol{\gamma_{12}} &= \label{eq:g12} \begin{aligned}[t]
	&-4 \ell^5 (x+1) \left(\frac{1}{\gamma_1}-\frac{1}{\gamma_2}\right)-4 \ell^4 (x+1)^2 \times\left(\frac{1}{\gamma_1}+\frac{1}{\gamma_2}\right) \\
	&+\ell^3 \Big(4 (x+1) (\gamma_1-\gamma_2) +2 (x+1) \left(\frac{1}{\gamma_1}-\frac{1}{\gamma_2}\right) \left(4 (x+1)^2+y^2\right)\Big) \\
	&+\ell^2 \Big(2 (\gamma_1+\gamma_2) \left(6 (x+1)^2+4 y^2\right) \\
	&\quad\,+2 \left(\frac{1}{\gamma_1}+\frac{1}{\gamma_2}\right) \Big(3 (x+1)^2 y^2 +4 (x+1)^4+y^4\Big)\Big) \\
	&+\ell \Big(-2 (x+1) (\gamma_1-\gamma_2) \left(2 (x+1)^2-4 y^2\right) \\
	&\quad\,-2 (x+1) \left(\frac{1}{\gamma_1}-\frac{1}{\gamma_2}\right) \Big((x+1)^2 y^2 +2 (x+1)^4-y^4\Big)\Big) \\
	&-2 \left(\frac{1}{\gamma_1}+\frac{1}{\gamma_2}\right) \left(2 (x+1)^2-y^2\right) \left((x+1)^2+y^2\right)^2 \\
	&-8 (\gamma_1+\gamma_2) \left(2 (x+1)^2-y^2\right) \left((x+1)^2+y^2\right) \\
	&+4 (\gamma_1+\gamma_2) \left((x+1)^2+y^2\right)^2 \end{aligned}
\end{align}

\clearpage

\end{appendices}


\bibliography{sn-article.bib}

\end{document}